\documentclass[aps,prb,twocolumn,superscriptaddress,longbibliography]{revtex4-2}

\usepackage[
  colorlinks=true,
  urlcolor=blue,
  linkcolor=blue,
  citecolor=blue,
  bookmarks=false,
  pdftitle={Graphene FQH phases Yukawa like},
  pdfauthor={Ankur Das}
]{hyperref}

\usepackage{graphicx}
\usepackage{amsmath}
\usepackage{amssymb}
\usepackage{float}
\usepackage[capitalise]{cleveref}
\usepackage{nicefrac}
\usepackage{multirow}
\usepackage{xcolor}
\usepackage{cleveref}

\usepackage{ytableau}
\usepackage{bm}
\usepackage{amsmath}
\usepackage{amssymb}
\usepackage{mathtools}
\usepackage{amsfonts}
\usepackage[all]{xy}
\usepackage{xspace}
\usepackage{tabularx}
\usepackage{caption, subcaption}
\usepackage{float}
\usepackage{tikz}
\usepackage{pgfplots}
\usepackage{pgfplotstable}

\pgfplotsset{compat=1.16, set layers, mark layer=axis tick labels}
\usepackage{multirow}
\usepackage{makecell}
\usepackage[normalem]{ulem}
\usepackage{pdfpages}

\usetikzlibrary{positioning}
\graphicspath{{FigurePRB/}}

\makeatletter
\AtBeginDocument{\let\LS@rot\@undefined}
\makeatother

\begin{document}
\title{Fractional Quantum Hall phases of graphene beyond ultra-short range intervalley-anisotropic interaction}

\author{Oleg Grigorev}
\affiliation{Department of Molecular Chemistry and Material Science, Weizmann Institute of Science, Rehovot 7610001, Israel}
\affiliation{Department of Condensed Matter Physics, Weizmann Institute of Science, Rehovot 7610001, Israel}
\author{Ankur Das}
\email{ankur@labs.iisertirupati.ac.in}
\affiliation{Department of Physics, Indian Institute of Science Education and Research (IISER) Tirupati, Tirupati 517619, India}
\affiliation{Department of Condensed Matter Physics, Weizmann Institute of Science, Rehovot 7610001, Israel}

\begin{abstract}
Recent experimental and theoretical development in the Quantum Hall effect in monolayer graphene
showed that the previous model of the valley-anisotropy interaction is incomplete, as it was
assumed to be ultra-short range (USR). In this work, we use exact diagonalization to go beyond the
ultra-short range to find the different phases for $\nu=2/3$. We model the interaction as Yukawa
so that we can control the range as a proof of concept. Even in this simple setting, we discovered how dropping the USR condition shifts the transition borders in favour of certain phases, leads to a new bond-ordered phase appearing, and breaks the ferromagnetic phase in two competing states as a result of lifting the USR-driven degeneracy.
\end{abstract}

\maketitle

\section{Introduction}

Quantum Hall effect is among the most cherished topics in condensed matter physics and has been thoroughly researched since its inception more than 40 years ago \cite{PrangeGirvin1990,dassarma1996:qhe,IQHE_Discovery_1980}. The reason for this is that it provides a viable platform for the study of strongly correlated electron systems, with Landau levels being naturally flat. For each Landau level, there is more than one flavour, e.g. spin, and only a few of them can be filled. Then, the ground state will be chosen by breaking these flavour symmetry due to interactions \cite{AliceFisher,YaDasSarmaMcD,Herbut2007_1,Herbut2007_2}.  It results in the phenomenon of quantum Hall ferromagnetism \cite{YaDasSarmaMcD,Nomura}. However, this is not restricted to the integer quantum Hall but is also applicable to the Fractional quantum Hall states. Even in fractional quantum Hall states, some of these internal symmetries will be broken \cite{IntiMacDonald,Wu}.

Observed as early as twenty years ago in multilayer heterostructures \cite{SchliemMacDonal}, these effects gained prominence with the rise of graphene and other sister materials, arguably because of sample preparation in graphene being relatively simpler and more independent experimental groups being now able to contribute to the expanding body of research. Modelled as Dirac electrons coming in two flavours (valley) with two spin polarizations, low-energy electrons in graphene provide a plethora of ground states at different fillings, even in monolayers.

A simple theoretical model of ultra short-range valley symmetry breaking electron-electron interactions was developed in Refs. \onlinecite{AliceFisher,Kharitonov}. Building on it, a number of emerging phases were theoretically predicted at charge neutrality in {monolayer graphene (MLG)} \cite{Kharitonov}. There have been a number of experiments that point towards these phases \cite{zhang2006:nu0,jiang2007:nu0,young2012:nu0,Maher_Kim_etal_2013} mainly targetted at the charge neutrality i.e. $\nu=0$. The canted antiferromagnet phase was observed in tilted field experiments \cite{Young} showing the transition to Ferromagnet. The transition to the ferromagnetic {(FM)} phase was reached in magnon transport experiments \cite{Young_magnon}. Bond-ordered K\'{e}kul\'{e} distortion {(KD)} state was observed by { scanning tunneling spectroscopy} \cite{Li,Coissard,Liu}, and further investigation led to discovering charge density wave {(CDW)} state \cite{Coissard,Liu}. However, more puzzling results were uncovered in STS as well, with an apparent CDW-KD coexistent phase not predicted in previous numerical studies \cite{Coissard}. Therefore, assumptions leading to the initial model were revisited, with claims being made that the ultra-short range (USR) approximation is too stringent. 

Some initial progress indicating the existence of new states in MLG at both charge neutrality and numerous fractional fillings was made in \cite*{Ank0,Ank1,StefInt,An,an2024fractionalquantumhallcoexistence} using Hartree-Fock mean-field theory. These do not make full use of the concrete form of non-USR interactions. One reason for that is that the exact expressions for this function remain an open question, though there were some recent developments in determining it in bilayer graphene \cite{KhannaZhu_2023}. We argue, however, that with the use of a model interaction, satisfying the requirements drawn of physical intuition, combined with exact diagonalization in a finite system, a number of conclusions can be drawn that will provide us with a number of qualitatively correct results which remained undiscovered. We chose Yukawa-like potential, naturally providing us with a range parameter and reducing to a USR interaction in a limit for such a model interaction.

The structure is as follows. In \cref{sec:valani}, we provide the necessary background on the model Hamiltonians for interacting electrons in graphene. In \cref{sec:EDInt}, we provide the results of exact diagonalization {(ED)}. We show that a new bond-ordered {(BO)} phase occurs in the case of both charge neutrality and fractional filling between FM and CDW states as the USR condition is dropped; in the case of fractional filling, we also show how the finite range lifts the added degeneracy in FM sector, leading to two competing ferromagnetic states, and shifts the transition of {antiferromagnetic (}AF{)} to KD state. In \cref{sec:disscussion}, we conclude by summing up the results and open questions. In appendices, we provide a concise calculation of all the elements needed for our exact diagonalization scheme, as well as some additional computational results.  

\section{Graphene with nonzero range valley symmetry breaking terms}
\label{sec:valani}

To compute the energy levels of electrons in monolayer graphene in a strong perpendicular magnetic field, we use the continuous model (Ref \cite{Castro}); we will discuss the adjustments we make to this model to take lattice-scale physics into account. The most simplistic picture is that of all the electron-electron interactions being ignored, as well as Zeeman and substrate-induced potentials (valley Zeeman) effects. Under such a condition, the emerging band structure can be approximated by Landau levels {(LLs)} for Dirac electrons ($E_n=\mathrm{sgn}(n) \sqrt{2 |n|} \frac{\hbar v_F}{l_B}\sim \mathrm{sgn}(n) \sqrt{|n| B}$) \cite{Castro}. Each level has an additional fourfold degeneracy due to electrons having different spin polarizations and valley-spin ``flavours"; in other words, all the states belonging to one LL lie in the same invariant subspace of $SU(4)$ spin-valley rotations.

If we do take electron-electron interactions into account, a natural division into magnetic length scale $SU(4)$ symmetric part (``long-range" Coulomb) and lattice-scale valley symmetry breaking terms naturally emerges \cite{AliceFisher}, as in a realistic magnetic field we have $\frac{l_B}{a}\sim10-100$ \cite{JungMc,Kharitonov}. What one would also expect is the cyclotron energy {$E_m=\frac{\hbar v_F}{l_B}$} exceeding Coulomb energy {$E_C=\frac{e^2}{\epsilon l_B}$}: {$E_m >> E_C$}. However, the ``graphene fine structure" constant $\frac{E_C}{E_m}=\frac{e^2}{\epsilon \hbar v_F}$ was shown to be in the range of $0.5-2.2$, depending on $\epsilon$ in a substrate of choice \cite{NayakPeters}.
Therefore, instead of being simplified via restriction to the 0LL, the Hamiltonian requires a tedious task of accounting for LL mixing by renormalizing electron interactions in the 0LL.

The first scheme was suggested by Kharitonov in Ref \cite{Kharitonov}. Building on the previous works (Refs \cite{Herbut2007_1,Herbut2007_2,AliceFisher,Alei2007}) and introducing several assumptions, namely, treating higher LLs as a continuum and explicitly demanding
all the symmetry-breaking terms to vanish at the magnetic scale, he arrived at the Hamiltonian 

\begin{eqnarray}\label{Cou}
H&=&H_C+H_v+H_Z,\\ H_C&=&\sum_{i<j} \frac{e^2}{\epsilon |r_{ij}|}, \\ H_v&=&\sum_{i<j} \left(g_z \tau^i_z \tau^j_z + g_{\perp}\left(\tau^i_x\tau^j_x + \tau^i_y \tau^j_y \right)\right) \delta^{(2)}(r_{ij}), \\ H_Z&=& - E_{Z} \sum_{i} \sigma^i_z
\end{eqnarray}

Here the $\tau^a$ and $\sigma^a$ stand for, respectively, spin and valley Pauli matrices. The degeneracy of the $SU(4)$ multiplet of pure Coulomb ground states is weakly lifted by valley $SU(2)$ symmetry-breaking terms. Therefore, despite symmetry breaking couplings $\frac{g_{\perp}}{2\pi l^2_B},\frac{g_z}{2\pi l^2_B}$ are far smaller than the Coulomb energy $\frac{e^2}{\epsilon l_B}$, they choose the ground state of Kharitonov's Hamiltonian. The phase diagrams of MLG with this model Hamiltonian were thus obtained in \cite{Kharitonov,IntiMacDonald,Wu,LeThierr} for charge neutrality case as well as for a number of fractional filling factors.

However, it was argued that the picture of ultra-short range intervalley interactions may be too simplistic \cite{Ank0,Ank1,StefInt} to explain the recent experimental findings \cite{Li,Liu, Coissard}; another renormalization group calculation performed in Ref \cite{Ank0} showed that the symmetry-breaking terms remain nonzero at magnetic-length scale. Said terms can now be expressed in the following form

\begin{equation}\label{sr}
H_v=\sum_{i<j} \left(g_z(r_{ij}) \tau^i_z \tau^j_z + g_{\perp}(r_{ij})\left(\tau^i_x\tau^j_x + \tau^i_y \tau^j_y \right)\right)
\end{equation}

Symmetry-breaking terms now depend on {unknown potential functions}. It proved difficult to compute using {renormalization group} techniques \cite{Ank0}, and without an exact expression for it we seem left with an infinite number of parameters governing the behaviour of MLG Hamiltonian (some progress has been made in bilayer graphene with experimental support \cite{KhannaZhu_2023}). There are several ways to reduce this uncertainty and take only essential parts of potentials into consideration.

\subsection{Hartree-Fock {(HF)} approximation}

Within variational approach, the ground state is a single Slater determinant (SSD), and only depends on a few numbers characterizing the interaction. No further specifics about the interaction are needed to assess the band structure of a strongly correlated system. 
This was done for half-filled zeroth Landau level in Refs \cite{Ank1,Ank0}. In this case, we can attribute two out of four occupied sublevels in 0LL to each candidate state and thus associate it with a projector $\Delta$ onto the occupied subspace. Energy of a state corresponding to $\Delta$ is then \cite{Ank1}

\begin{align}
    \frac{\mathcal{E}_{HF}}{N_{\phi}} = \frac{1}{2}\sum_{a=x,y,z} g_{a,H}\left({\mathrm Tr} (\tau_a \Delta)\right)^2 - g_{a,F} {\mathrm Tr} \left(\left(\tau_a \Delta\right)^2\right) 
\end{align}
where, if $\tilde{V}(q)$ stands for a Fourier transform of pair potential, $g_{a,H}=\frac{1}{2\pi l^2_b}\tilde{V_a}(0)$ and $g_{a,F}=\frac{1}{{(2\pi)}^2} \int_{\mathbb R^2} \tilde{V_a}(q)e^{-\frac{q^2l^2_B}{2}}d^2 q$ are Hartree (direct) and Fock (exchange) couplings. This expands on the expression used in Ref. \onlinecite{Kharitonov} for pointlike potentials, for which  $V_{H}=V_{F}$. For an expression where this identity no longer holds, it proved that states delivering extrema do not have to be either spin or valley ordered. Elaborated further, this approach led to finding evidence of new phases as well as their coexistence with those already discovered in the charge neutrality case. There were recent reports of using this method for certain fractional fillings as well \cite{An}. 

However, the state remaining a pure SSD is still limiting and prevents finding regions of parameters where a true ground state is more complicated. Besides, accounting for candidate states given by polarizations of their flavour components is rather involved even for the case of charge neutrality and becomes even more tedious for fractional fillings. In this paper, we will follow a more impartial method of exact diagonalization, which will also allow us to assess the low-lying excitations. However, other simplifications are due to make the problem tractable.

\subsection{Model pair potential and pseudopotentials}

The influence of the potential on the band structure of the system under consideration still boils down to its dependence on a set of numbers; they are known as Haldane pseudopotentials (Ref. \onlinecite{Haldane89,Haldane83}). We can, therefore, encode all the physically relevant features in a certain parametric family of model potentials. 

Pseudopotential $V_n=V(n\hbar)$ can be thought of as the interaction energy of two electrons with relative angular momentum $n\hbar$. For USR interaction, all the $V_{>0}=0$, so $V_0$ is the only factor at play. One can consider changes in the features the system exhibits with the interaction obtaining range as the perturbation caused by subleading pseudopotentials moving away from zero.

The natural choice of the parameter we add to tweak the interaction is its range; we want the model pair potential $g_0 u_{\alpha}(r_{12})$ to reduce to USR limit with $\alpha$ (inverse range) tending to infinity. The range gives a scale at which the interaction is still non-negligible but decreases outside of it. The simplest case is when the electrons with close (but not equal) angular momenta become coupled, with coupling weakening with larger relative angular momentum.

Lead by this logic, we picked ``Yukawa-like" pair potential $V(r)=\frac{g_0 \alpha}{2\pi}\frac{e^{-\alpha r}}{r}$ for a model interaction. Yukawa pseudopotentials $V_n$ decrease monotonously with $n$. Pushing $\alpha\rightarrow \infty$ gives the USR interaction $V(r)=g_0 \delta^{(2)}(r)$. Curiously, you can think of this interaction modelling the lattice-scale part of Coulomb as a ``screened Coulomb" with a ``Debye radius" depending on direction.

{Allowing us to investigate the effects of range, this potential has a shortcoming. For a realistic model, it is expected that the interaction would oscillate between attractive and repulsive with distance \cite{KhannaZhu_2023, Ank0, Ank1, Goerbig1,Goerbig10}. In terms of pseudopotentials, $V_n$ are then shifting their signs. To include this into our analysis, we also considered ``oscillatory Yukawa" potential $V(r)=\frac{g_0 \alpha (k^2+1) }{2\pi}\frac{e^{-\alpha r}\cos(k\alpha r)}{r}$. The normalization is chosen to achieve the correct factor in the USR limit. For our study, expecting the rate of oscillation to be faster than the rate of decay, we chose $k=2$ for simplicity.}

It is also noteworthy that one can deduce the Hartree and Fock couplings from $V_n$ (the details are given in Appendix \ref{pseudopo})  and, therefore, collate the numerical results obtained in this work to those done by variational methods {(see Appendix \ref{dougandvladim})}.
`
\section{Exact diagonalization  scheme and results}
\label{sec:EDInt}
It is convenient to rewrite the symmetry-broken Hamiltonian (\ref{sr}) in terms of fermion operators (we will ignore Zeeman and valley-Zeeman terms unless stated otherwise in the rest of the paper to make the way pair potential influences the energy levels clear)
\begin{multline}\label{H}
H=\sum_{I_1,I_2,I'_1,I'_2} \left(V^{C}_{I_1,I_2;I'_1,I'_2}+\right. \\ \left. g\cdot V^{SR}_{I_1,I_2;I'_1,I'_2}(\theta, \alpha_{\perp}, \alpha_z)\right) c^{\dagger}_{I_1}c^{\dagger}_{I_2}c_{I'_1}c_{I'_2}
\end{multline}
where $I_j$ runs over all combinations of available one-particle angular momenta, valley, and spin quantum numbers. The four-fermion coefficients are sparse and possess numerous symmetries; for the exact expression as well as the details of the calculation, see \cref{hamdige,matelem}. Here, we point out their most important qualitative features.

The coefficients are sums of Coulomb term and valley {(oscillatory)} Yukawa; the latter depends on the ranges and projectively depends on the strengths of interaction. That is, we can express it as a function of {$\theta=\arg (g_z+ig_{\perp})$}, with dependence on the radial term $g=\sqrt{g^2_z+g^2_{\perp}}$ brought down to a mere prefactor. 

In the absence of symmetry-breaking terms, the Hamiltonian has full $SU(4)$ symmetry. The ground eigenspace, therefore, is an irreducible representation of $SU(4)$ or a multiplet. To each of those, one may assign the ``flavour filling factor" in the following way. Suppose we fix a direction of polarization $z$ and define the generators of $SU(4)$ w.r.t. this choice. Then, each multiplet, similarly to the case of $SU(2)$, has a number of highest-weight vectors. For each of those, the flavour ($\uparrow K$, $\uparrow K'$, $\downarrow K$, $\downarrow K'$) population are good quantum numbers; the ordered quadruple of them fully defines the multiplet. (see Appendix \ref{trunca} for more details {as well as generators of $SU(4)$ given in fermionic terms)

As the energy scale of symmetry-breaking terms is nonzero yet small compared to Coulomb, the band structure of our system is a series of levels perturbatively fanning out of consecutive $SU(4)$ multiples, with leftover $SU(2)_s\times U(1)_v \times (\mathbb{Z}_2)_v$ symmetry ($s$ subscript stands for spin and $v$ for valley). As the Coulomb term contribution to each state descending from a multiplet is roughly the same, we can take all parameter dependencies of $V^{SR}$ into account by taking fixed ranges $\alpha_{z,\perp}$ and plotting the spectrum depending on $\theta$.

The Hilbert space we use is that of $N_e$ electrons staying in the 0LL with arbitrary spin and valley polarization and the angular momentum not larger than $N_\phi \hbar$ -- therefore, occupying $4N_{\phi}$ available sites. In other words, we perform exact diagonalization in disk geometry. The Hamiltonian (\ref{H}) commutes with the net spin $S$, net valley polarization $T_z$ and net angular momentum operators; therefore, Hilbert space is divided into sectors where we can perform calculations independently, and we can assign relevant quantum numbers to candidate states. {As we are interested in the sector where the ground state lies, and it is expected to be uniform in space, we can choose the corresponding angular momentum sector. This both decreases the computational cost and allows us to study only momentum-conserving excitations that are due to $SU(4)$ symmetry breaking.  }

Disk geometry is usually not the first choice for quantum Hall systems. Although
it was introduced in the pioneering paper on the topic \cite{Laughlin-83}, it demonstrates poor performance in systems with long-range interactions \cite{KasnerApel}.
However, the long-range part -- Coulomb coupling -- gives us the base level, and correction to it is what we are most curious about. The interaction causing these corrections, on the other hand, decreases exponentially outside the region limited by magnetic length or less; short-range {(SR)} interactions like that match the technique much better \cite{KasnerApel}. 
 
On the other hand, two alternative geometries, on a sphere \cite{Haldane83} and on a torus \cite{Haldane86,Yoshioka}, have their own shortcomings.
In spherical geometry for a multicomponent system, the Wen-Zee shift \cite{WenZee} (filling factor differing from $\frac{N_e}{N_{phi}}$) leads to a following discrepancy. If we have two systems with the same combined number of electrons but different flavour contents, we end up with different filling factors; this adds an artificial complication to comparing the band structures of those systems.

Torus geometry adds a topological degeneracy to the energy levels for certain filling factors. It obscures the symmetries of the ground states, complicating the job of matching the ED results with the corresponding variational theory predictions. However, perhaps more importantly, it bloats the Hilbert space, requiring more computational resources to process.
Taking these considerations into account, we choose disk geometry as both more natural and less computationally demanding and robust enough for the sort of interaction we are studying. {We used sphere geometry as a secondary choice to cross-check our results for charge neutrality, see Appendix \ref{orb}.}

\subsection{Charge neutrality}

\begin{figure*}
 \includegraphics[angle=270,width=\linewidth]{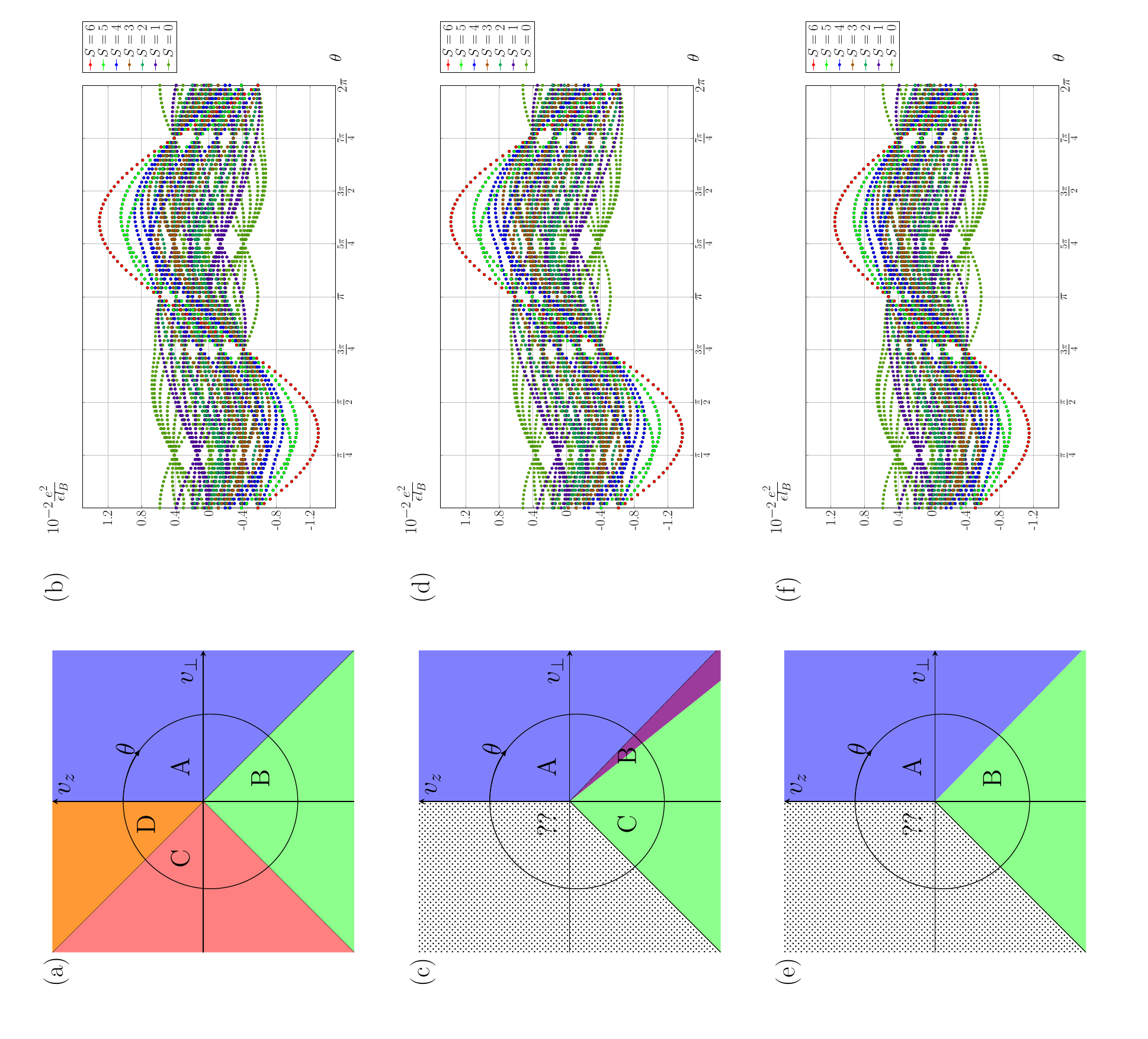}
	\caption{\small{Energy spectra for interacting electrons in graphene at charge neutrality, $E_Z=0$, $N_{\phi}=6,~N_e=12$. The Coulomb-favoured $SU(4)$ (6,6,0) multiplet is broken by weak valley {(oscillatory)} Yukawa coupling.  Corrections to Coulomb energy against the interaction anisotropy {$\theta=\arg \left( u^{z}_0+iu^{\perp}_0\right)$} (b, {d}{, f}) and proposed phase diagrams (a, c{, e}}) are shown. States of different spins are highlighted in different colours.
    (a, b) The `almost USR' interaction, $\alpha^{-1}_{\perp,z}=10^{-4}l_B$. 
    (c, d) {The oscillatory Yukawa interaction, $\alpha^{-1}_{\perp}=\alpha^{-1}_{z}=l_B/2$. (e,f)} The {Yukawa} interaction, $\alpha^{-1}_{\perp}=\alpha^{-1}_{z}=l_B/3$.
	}	\label{halffill}
\end{figure*}

We start by presenting and analyzing the results of our calculations for the case of charge neutrality. We took $N_e=12,~N_{\phi}=6$ (thus two out of four spin-valley sublevels of 0LL filled). We take the square mean Yukawa coupling $\frac{g}{2\pi l^2_B}=0.0{0}1 \frac{e^2}{\epsilon l_B}$. {The ground state multiplet of a symmetry-unbroken system lies in the sector $L=30\hbar$, as the expected  uniform population of momentum sites gives $L=2\sum_{n=0}^{N_{\phi}-1}n\hbar=30\hbar$. We confine our study to this sector}. \cref{halffill} shows the (6,6,0) multiplet for cases of almost ultrashort and short-range interactions.

\subsubsection{USR $\alpha_{\perp}=\alpha_z=(10^{-4} l_B)^{-1}$}

{$0<\theta<\frac{3\pi}{4}$}: the ground state belongs to the $T_z=0$, $S=6$ sector. Having the highest possible spin, this state can be identified as ferromagnetic. Due to unbroken $SU(2)_{s}$ symmetry, it is natural that the ground state has $2S+1=13$-fold degeneracy (therefore, the eigenspace comprises the whole $T_z=0$, $S=6$ sector, and the base states are pure Slaters). These results are consistent with both the mean-field approach in ref. \onlinecite{Kharitonov} and ED results in toric geometry ref. \onlinecite{Wu}.

{$\frac{3\pi}{4}<\theta<\frac{5\pi}{4}$}, the state comes from  $S=0$, $T_z=\pm 6$ sector (double degenerate with $(Z_2)_v$ acting on the eigenspace, base states of which are again pure Slaters). This we identify as a valley ferromagnet (CDW state), also consistent with \cite{Kharitonov,Wu}. The transition it undergoes from FM to CDW is first-order, as indicated by a true level crossing.

{$\frac{5\pi}{4}<\theta<2\pi$}, the ground state is from $T_z=S=0$ sector with no degeneracy. The quantum numbers are consistent with both $KD$ and $AF$ states predicted by mean-field techniques; however, there are some key differences. First of all, the eigenstates predicted by our ED analysis are mixed. Second, degeneracy predicted by HF studies is no longer present in a finite system; e.g., AF state, having the largest N\'{e}el vector, cannot be a singlet. With this in mind, it is not surprising that ground state level crossing is absent at $\frac{7\pi}{4}$, where a BO-AF transition was predicted in \cite{Kharitonov}. A level caving in towards the crossing of the first excited levels, however, hints at a possible tower of state collapse \cite{EDquirk} as the finite system degeneracy lifting begins to wane as the system size grows. As this particular calculation was more meant to align our results with ones already established, we refrained from further analysis. It should be pointed out, though, that in ref. \onlinecite{Wu}, a similar feature was observed in toric geometry.

The results for USR regime largely agree with preexisting findings. With this additional validation, we apply ED with disk geometry to the more intriguing case of short-range Yukawa interactions.

\subsubsection{{Oscillatory Yukawa $\alpha_{\perp}=\alpha_z=(1/2 l_B)^{-1}$}}

The ferromagnetic sector ($T_z=0,~S=6$) occupies a smaller sector {$0<\theta<\frac{3\pi}{4}-\delta$}, {$\delta\approx 0.012\pi$}.

On the border of the ferromagnetic sector and CDW, a new phase appears, with $S=0,~T_z=0$. For the parameters we chose, it spans roughly the slice {$\frac{3\pi}{4}-\delta<\theta<0.7875\pi$}. We point out that it shares the spin-valley subspace with BO states, which is in line with lattice calculations in Ref \cite{Ank0}, showing that with symmetry-breaking interaction gaining range, a number of BO regions emerge. As the range becomes wider, the new phase domain increases, though its dependence on the interplay of different ranges is awaiting further research.

The CDW area ($S=0,~|T_z|=6$) is smaller, occupying the sector {$0.7875{\pi}<\theta<\frac{5\pi}{4}$}, as is the energy gap, while the rightmost transition point to the valley unpolarized state remains unaffected.

As before, in the region {$\frac{5\pi}{4}<\theta<2\pi$} there is no level crossing. The ground state has no degeneracy and has $T_z=S=0$. It is tempting to treat the first excitation crossing as another indication of BO-AF transition happening at the thermodynamic limit; however, this question deserves a more careful treatment {(see also Appendix \ref{dougandvladim})}, and we will address it elsewhere.

A new phase appearing at the border of FM and CDW alone is unambiguous evidence that the interaction gaining nonzero range entails novel features in the phase diagram for charge neutrality. We mention in passing that the gaps between the ground state and low excited ones look largely unaffected in ``spin-ordered" zones and markedly different in ``valley-ordered" parts. This is of interest for further inspection of the excitation structure.

\subsubsection{{Yukawa} $\alpha_{\perp}=\alpha_z = (1/3 l_B)^{-1}$}

{Compared to the USR case, no new phases are observed. The FM sector ($T_z=0,~S=6$) is again decreased, with its border to CDW state now moved to $\theta=0.7399 \pi$. The CDW sector then continues up to $\theta=\frac{5\pi}{4}$. 

Similar to both former cases, there is no level crossing in the region $\frac{5\pi}{4} < \theta < 2\pi$. The first excitation crossing, curiously, shifts clockwise, in the opposing direction to that of oscillatory Yukawa.}

\subsection{Filling $\frac{2}{3}$}\label{frf}

We studied the way energy levels are affected by long-range interaction for a certain filling factor $2/3$. We chose it as it demonstrates additional degeneracy, which, as we show below, is lifted by relaxing the USR assumption.

The case of fractional 0LL fillings was extensively treated in Ref \cite{IntiMacDonald} using an extension of HF method. In Ref \cite{LeThierr}, ED calculations were performed for a number of fractions in the USR limit, validating some mean-field results and contradicting others. One of the key takeaways of that research was that the picture of phases for $\nu$ fillings is interlinked with the one emerging for charge neutrality. The nomenclature of these fractional phases is, therefore, defined by the state of underlying filled shells. {The (6,6,4) multiplet we considered lies in the angular momentum sector $40\hbar$. This value can be defined, as previously, by applying uniformity condition, averaging the momentum for all the possible electron configurations.}

\begin{figure*}
 \includegraphics[angle=270, width=\textwidth]{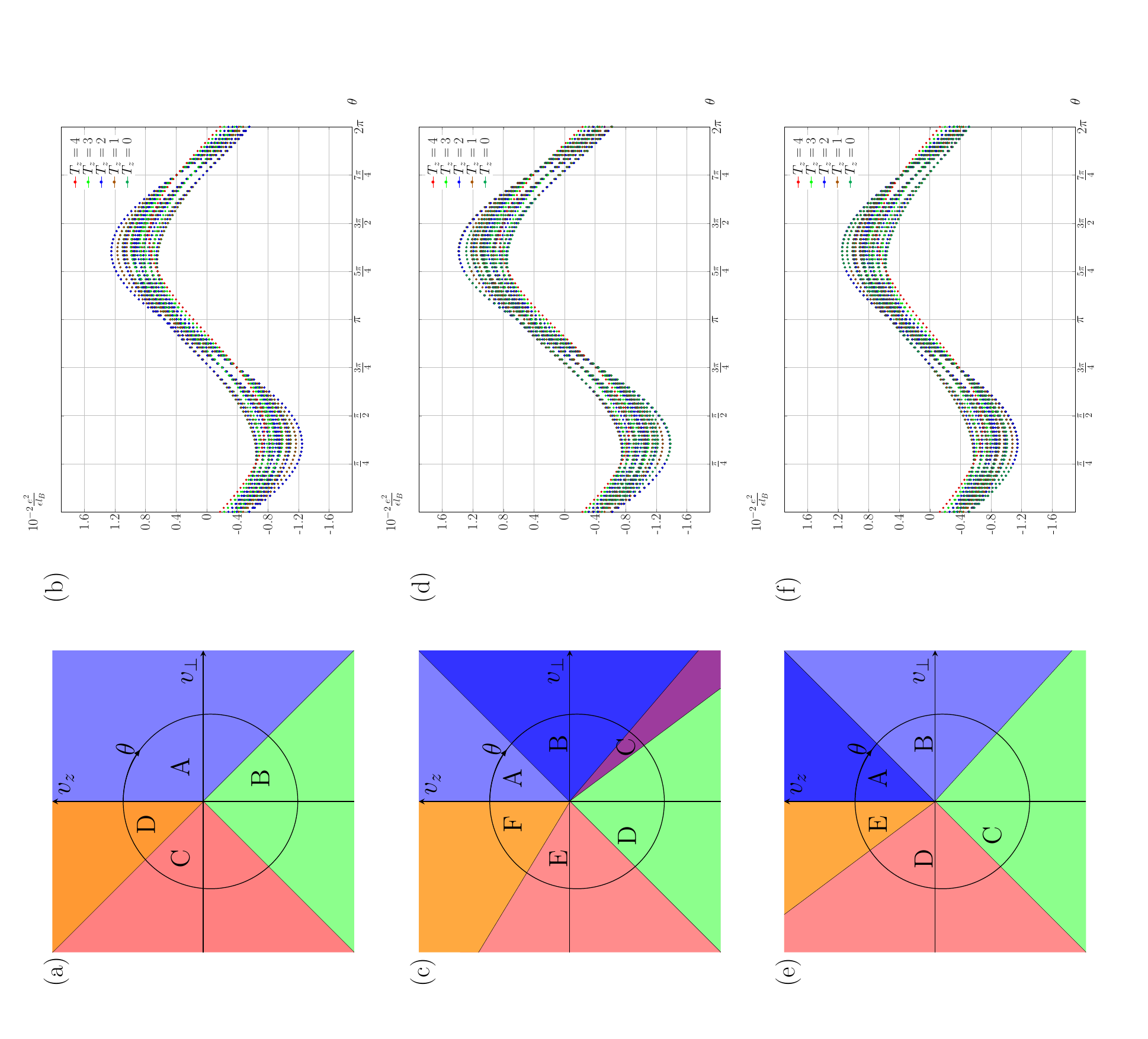}

	\caption{\small{Energy spectra for interacting electrons in graphene at filling factor $2/3$, $E_Z=0$, $N_{\phi}=6,~N_e=16$. The $SU(4)$ (6,6,4) multiplet is broken by weak valley {(oscillatory)} Yukawa coupling.  Corrections to Coulomb energy against the intervalley anisotropy {$\theta=\arg \left (u^{z}_0+iu^{\perp}_0\right)$} (b, d{, f}) and suggested phase diagrams (a, c{, e}) shown. Different valley polarizations are shown in different colours.
			(a, b) The USR interaction,  $\alpha^{-1}_{\perp,z}=10^{-4}l_B$. 
			(c, d) The {oscillatory Yukawa} interaction, {$\alpha^{-1}_{\perp}=\alpha^{-1}_{z}=l_B/2$}. {(e, f) Yukawa interaction, $\alpha_{\perp}^{-1}=\alpha_z^{-1}=l_B/3$}
	}}	\label{fig:f}
\end{figure*}

\subsubsection{USR}

{$0<\theta<\frac{3\pi}{4}$} -- the states with valley polarization $T_z=[-2;2]$ and spin $S=4$. The spin polarization is the highest possible for a state in the chosen multiplet, so we identify those states as ferromagnetic. Despite the Hamiltonian lacking $SU(2)_v$ symmetry, both $SU(2)_v$ and $SU(2)_s$ act nontrivially on this eigenspace, with the presence of valley degeneracy due to the interactions being pointlike. This phenomenon has already been addressed in Ref. \onlinecite{LeThierr} for another filling factor; we take a somewhat different angle, which will be further unravelled in the next subsection. Atop the ferromagnetic shell, we have four more electrons, all occupying the same valley flavour; due to the Pauli principle, they all have different angular momenta. As the only pseudopotential that is nonzero is $V_0$, it has no effect on the energy of the states. This, however, ceases to be true for the short-range $V$ we clarify below.

{$\frac{3\pi}{4}<\theta<\frac{5\pi}{4}$} -- states with valley polarization at its highest and $S=2$. Built atop the completely valley-polarized shell, those states represent the CDW phase.

{$\frac{5\pi}{4}<\theta<\frac{7\pi}{4}$} -- here the ground level has quantum numbers $T_z=0,S=2$ (a spin multiplet). The region and quantum numbers are in accord with \cite{IntiMacDonald}, so we identify this phase as K\'{e}kul\'{e} distortion.

{$\frac{7\pi}{4}<\theta<2\pi$} -- the state with quantum numbers $T_z=2,S=2$, which we identify as antiferromagnetic. Unlike charge neutrality case, the bond ordered and AF states now belong to different spin-valley sectors, therefore transition between them now should include a level crossing; hereinafter we can safely make a conclusion that a first-order transition happens.

As was already mentioned, the fractional phases are ``built atop" the charge neutrality counterparts, as illustrated by the phase diagram, summed up in Fig. \ref{fig:f}. There, the partition into phases (and, to an extent, the nature of these phases) is the same as in the charge neutrality case. However, as we introduce corrections to the USR Hamiltonian, the phase diagrams diverge. In \cite{IntiMacDonald}, the corrections caused by Zeeman splitting were researched; we show how introducing range affects the picture.

\begin{figure*}
     \includegraphics[angle=270,clip,trim=0cm 0cm 0cm 7cm,width=\linewidth]{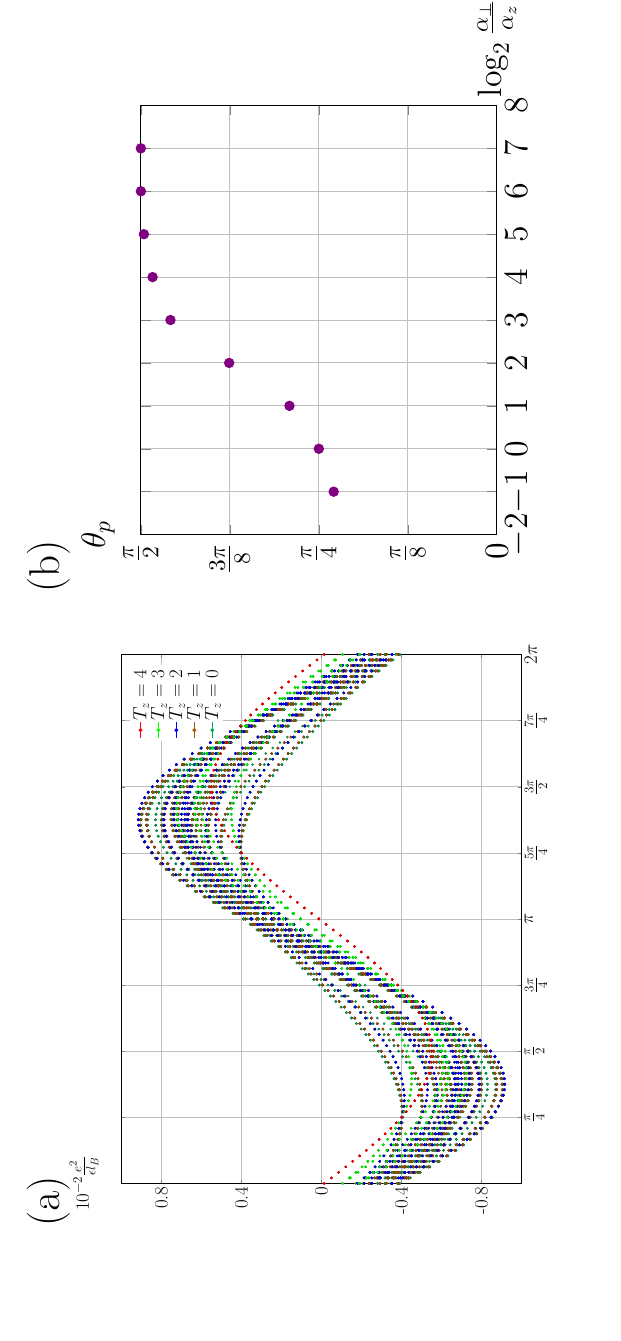}
    \caption{(a) Energy spectrum for interacting electrons in graphene at {filling fraction} 2/3, $E_Z=0,~N_{\phi}=6,~N_e=16$; Yukawa interaction parameters $\alpha_{z}l_B=1,~\alpha_{\perp}l_B=1.3$, (6,6,4) $SU(4)$ multiplet shown. (b) In the same type of system, $\alpha_z l_B$ is fixed and equal to 2, while $\alpha_{\perp} l_B$ changes. The angle $\theta_{FM}$ where Ising-FM -- XY-FM transition happens is plotted against $\log_{2}\frac{\alpha_\perp}{\alpha_z}$}\label{swap}
\end{figure*}

\subsubsection{{Oscillatory Yukawa ($\alpha_{\perp z}^{-1}=l_B/2$)}}

The $V_{>0}$ pseudopotentials no longer vanish; the degeneracy we mentioned in the previous section is now lifted. We have ``Ising-like" ferromagnet from $0$ to $\frac{\pi}{4}$, with $S=4,~|T_z|=2$ and XY-like ferromagnet with $S=4,~T_z=0$ in the rest of the ferromagnetic sector. The transition is first-order.

As in the half-filled system, a new state arises at the border of the ferromagnetic and CDW sectors. {It occupies a sector reducing FM by approximately $0.031\pi$ and CDW by $0.053\pi$}. The {spin of the ground state in this sector stays equal to $S=2$, while the projection of valley $|T_z|$ stays equal to 0 in a large part of the sector then gradually changes to 3 in a narrow strip adjacent to the CDW sector. This might be an indication of a continuous transition in the thermodynamical limit.}
Similarly to the USR interaction, the level crossing happens at $\frac{5\pi}{4}$, indicating a first-order CDW-BO transition. 
The KD sector ($S=2,T_z=0$) loses ground to the AF sector ($S=2, T_z=2$). The transition happens { at $\theta\approx 1.6437 \pi$}.

\subsubsection{{Yukawa ($\alpha^{-1}_{\perp z}=l_B/3$)}}

{Nonzero higher pseudopotentials again lift the degeneracy in ferromagnetic sector. However, here the XY-like ferromagnet occupies the sector $0<\theta<\frac{\pi}{4}$, followed by transition to Ising like FM state.

Similarly to what we saw in charge neutrality case, other differences from the USR phase diagram are confined to the area along the line $g_z+g_{\perp}=0$, with no other added phases, with (Ising) FM-CDW transition shifting counterclockwise 
and  KD-AF transition shifting clockwise
}

As we can see, with our model intervalley interactions gaining range, several novel features appear in the phase diagram. First, two competing ferromagnetic states that were indistinguishable at the USR limit emerge; the point where the spin-flop transition between them happens depends on the interplay of ranges. We elaborate on it further in  \ref{ising}. We also provide the band structures for larger ranges

Second, similarly to the charge neutrality case, there are some drastic changes along the $g_z+g_{\perp}=0$ line, where in the USR limit, the system demonstrates a high-level $SO(5)$ symmetry. We can state for certain that a new phase appears between the FM and CDW {for oscillatory short range interactions}; numerical experiments suggest that it becomes more prominent as the ranges become longer. The border between AF and KD demonstrates a very curious behaviour as well, shifting and leaving the nature of AF-KD transition an open question; interestingly, there were suggestions of exotic transition between these phases in bilayer graphene \cite{WZW}. These features deserve a separate detailed assessment, which is beyond the scope of this work.

{
\subsection{Spin flop transition}\label{ising}

Ferromagnetic phase splitting into two competing versions in fractional filling $\frac{2}{3}$ with interactions gaining range is potentially a useful feature that can be tested in experimental settings. The presence of a phase transition between two distinct ferromagnetics may prove that effective interactions have range, and its details may provide a glimpse into the specifics of these interactions. One of the simplest properties that can be additionally studied is how the predominance of each of these phases changes when the ranges are not isotropic.

{To showcase the phase diagram}, we chose an isotropic system because, not contradictory to our expectations, the phase transition occurred at line $v_z=v_{\perp}$. As a level crossing occurs in one of the test points, the transition is more evident for this system. But it is interesting to study the dynamics of this transition with changing ratios of ranges. 

For this study, we chose {Yukawa interaction with} a slightly larger range than the one considered in {\ref{frf}}. This way, although it makes the whole phase diagram harder to map, the transition we are focusing on became clearer (cf left panel of \cref{swap}, for $\alpha_z l_B=1$, $\alpha_{\perp} l_B=1.3$, with degeneracy lifting clearly visible on a chart). We fix $\alpha_z l_B = 2$ and give different values to $\alpha_{\perp} l_B$, going through the powers of 2. This way, we can access the asymptotic behaviour as $\alpha_{\perp}l_B$ goes to the USR range. To put a picture in a broader frame, we also included systems with $\perp$ range larger than $z$ range; these systems being less suitable for disk geometries, we considered only a couple of such terms.

We see that as the $(\alpha l_B)^{-1}$ range tends to zero, the transition moves in favour of {$XY$} ferromagnetic phase, swapping the whole quadrant as valley-valley interaction stays finite range in $z$ direction but becomes USR in perpendicular. Other numerical experiments do not indicate that this limit depends on $\alpha_z$; thus, a region of the phase diagram where {Ising-}FM phase persists remains even if only one component of interaction is non-USR.
}

\section{Discussion and summary}

\label{sec:disscussion}

In this work, we show a proof of concept in understanding the effect of anisotropy interactions beyond the USR assumptions. We considered the case of neutrality and filling factor $\nu=\frac{2}{3}$. For both of these cases, the phase diagram in the USR limit and without Zeeman splitting includes four phases, those being antiferromagnetic, K\'{e}kul\'{e} bond ordered, ferromagnetic and charge density wave. These results are in agreement with the previous studies.
Presenting hardcore interaction as a marginal case of a one-parametric family of interactions modelled by {(oscillatory)} Yukawa potential, we moved away from it by changing ranges. For the system thus obtained, we computed the energy levels using exact diagonalization in disk geometry. The structure of the latter showed how the simple four-phase diagram changes with the appearance of the range.

At charge neutrality, we saw curious dynamics along $g_z+g_{\perp}=0$ high symmetry line. Along the border of F and CDW phases, a new bond ordered phase occurs {in case of oscillatory potential}. The transition between AF and KD phases shifts in favour of {AF for oscillatory and KD for regular Yukawa interaction}. The borders of AF-F and CDW-KD remain unchanged, with level crossings indicating first-order transitions.

For the case of $\frac{2}{3}$, the picture is expected to repeat some dynamics of the underlying half-filled system; indeed, we saw a shift in AF-KD transition and a new bond-ordered phase between F and CDW. The quantum numbers inherent to AF and KD phases in the system are different, unlike the neutrality case, and thus, we were able to witness a level crossing, making more convincing evidence of a phase transition; the nature of this transition, though, remains an open question.
Another finding that was special for this particular filling factor was the observation of two different ferromagnetic phases emerging, with our model interactions lifting a degeneracy specific to the pointlike interaction in this fraction. This may be a promising way to test the validity of our model in experiments. Unlike the new bond-ordered phase we mentioned, those ones are expected to present at a wide range of parameters, even with interaction slightly deviating from the USR one.

The accurate model for intervalley interaction, to the best of our knowledge, is still not available. Instead, we chose the simplest ansatz demonstrating physically justified properties to make initial predictions of emerging phases that can be proved or disproved experimentally in the future. Curiously, some of our predictions align with results obtained by completely different methods \cite{Ank1, WZW}. We intend to conduct a more detailed study of the AF-KD transition and investigate the connection between breaking $SO(5)$ symmetry and non-USR potentials. We hope this work serves as proof of concept for researching more involved systems. 
This can be extended to other interesting fractions, such as double-layer graphene, bilayer and trilayer graphene, and non-abelian phases. We would also like to do some future studies where we would like to change externally controlled parameters like spin and valley Zeeman.

\begin{acknowledgments}
We thank Yuval Gefen and Ganpathy Murthy for useful discussions. O.G. gratefully acknowledges the support and hospitality of Weizmann Institute of Science, emergency program coordinator Joel Sussman and his host during his stay at the Condensed Matter Department, Yuval Gefen, among others. A.D. was supported by DFG MI 658/10-2 and DFG RO 2247/11-1. A.D. also thanks the Israel Planning and Budgeting Committee (PBC) and the Weizmann Institute of Science, the Dean of Faculty fellowship, and the Koshland Foundation for financial support, as well as the IISER Tirupati start-up grant. Our implementation of the exact diagonalization method was based on the open-source DiagHam package.
\end{acknowledgments}

\appendix

\section{Hamiltonian for disk geometry}\label{hamdige}

Making the representation of the interaction parts of Hamiltonian $H_C+H_v$ (\ref{Cou}),(\ref{sr}) more appropriate to use in ED, we rewrite it in second-quantized language

\begin{widetext}

\begin{equation}\label{sq}
H_{SR}=\frac{1}{2} \underbrace{\int d^2 r_1 d^2 r_2 V_0(r_{12}) N(r_1) N(r_2)}_{H^0_{SR}}+ \frac{1}{2} \underbrace{\int d^2 r_1 d^2 r_2 \left(V_{\perp}(r_{12}) \left[T_x(r_1)\cdot T_x(r_2)+T_y(r_1)\cdot T_y(r_2)\right]+ V_z(r_{12})T_z(r_1)\cdot T_z(r_2)\right)}_{H^{\perp}_{SR}+H^z_{SR}},
\end{equation}
\end{widetext}
where $V_0(r)=\frac{e^2}{\epsilon r}$ is Coulomb and $V_{z,\perp}=v_0\alpha_{z,\perp} \frac{e^{-\alpha_{z,\perp} r}}{r}$ {$\times (k^2+1)\cos(k \alpha r),k=0,2$} are Yukawa-like potentials, $N(r)$ is the occupation number operator, and $T_a(r)$ are second-quantized valley operators, which are easily expressed through field operators as

\begin{eqnarray}
N(r)&=&\sum_{t=K,K'}\sum_{s=\uparrow,\downarrow} \hat{\psi}^{\dagger}_{t,s} \hat{\psi}_{t,s} \\
T_a(r)&=&\sum_{t_1,t_2}\sum_{s}\hat{\psi}^{\dagger}_{s t_1}(r)\left(\tau_a\right)_{t_1 t_2}\hat{\psi}_{s t_2}(r)
\end{eqnarray}
with field operators being expressed via 0LL wavefunctions

\begin{equation}
\hat{\psi}_{s,t} (r) = \sum_{j=0}^{N_{\phi}-1} \left<r\right| \left. 0, j \right>_{s,t} \hat{a}_{j,s,t}
\end{equation}

Plugging in all of the above into (\ref{sq}), and with some simple Pauli matrix algebra, we obtain ($j_k$ run over all possible angular momenta, $s_l$ run over all possible spin polarizations)
\begin{widetext}
\begin{eqnarray}\label{h_unf}
H^0_{SR}&=&\sum_{s_k,t_k} \sum_{j_1,j_2,j_3,j_4} V^{0}_{j_1j_2j_3j_4} a^{\dagger}_{j_1,s_1,t_1}a^{\dagger}_{j_2,s_2,t_2}a_{j_3,s_2,t_2}a_{j_4,s_1,t_1}; \\
H^{\perp}_{SR}&=& \sum_{s_1,s_2}\sum_{j_1,j_2,j_3,j_4} 2 V^{\perp}_{j_1j_2j_3j_4} \left(a^{\dagger}_{j_1,s_1,K}a^{\dagger}_{j_2,s_2,K'}a_{j_3,s_2,K}a_{j_4,s_1,K'} + a^{\dagger}_{j_1,s_1,K'}a^{\dagger}_{j_2,s_2,K}a_{j_3,s_2,K'}a_{j_4,s_1,K}\right); \\
H^z_{S-R}&=& \sum_{s_1,s_2}\sum_{j_1,j_2,j_3,j_4}  V^{z}_{j_1j_2j_3j_4} \left(a^{\dagger}_{j_1,s_1,K}a^{\dagger}_{j_2,s_2,K}a_{j_3,s_2,K}a_{j_4,s_1,K} + a^{\dagger}_{j_1,s_1,K'}a^{\dagger}_{j_2,s_2,K'}a_{j_3,a_2,K'}a_{j_4,s_1,K'}\right).
\end{eqnarray}
\end{widetext}
with the four index coefficients being expressed through the coordinate part of 0LL states 

\begin{equation}
    V^{\cdot}_{j_1j_2j_3j_4}= \left< 0, j_1; 0, j_2\right| V^{\cdot} \left| 0, j_3; 0, j_4\right> 
\end{equation}

\section{Matrix elements}\label{matelem}

In this work, we study how the energy levels of a system depend on a certain set of parameters (in our case, ranges of intervalley interactions) by tweaking which we change the pair potential. To make the calculations of matrix elements more effective, it is useful to single out those parts that essentially depend on these parameters and those that are purely geometric.
The former is limited to a series of numbers called Haldane pseudopotentials \cite{Haldane83}, already mentioned in the main text.

We deal with a system of charged particles subject to a magnetic field, interacting via rotationally invariant pair potential $V_{\alpha}(r_i,r_j)=V_{\alpha}(|r_i-r_j|)${, where $r_i=\left(x_i\atop y_i\right)$}.
Limiting ourselves to a projection of $V$ to 0LL, and, if $M$ stands for the center of mass and $m$ for relative angular momentum:

\begin{equation}
\hat{V}=\sum_{M,m}\left<M,m\right|V\left|M,m\right>\left|M,m\right>\left<M,m\right|=\sum_{m} V_m \hat{P}_m
\end{equation}

Haldane's key observation was that $V$ matrix elements $V_m=\left<M,m\right|V\left|M,m\right>$ only depend on $m$, justifying the last equation. $V_m$, called Haldane pseudopotentials encode the essential part of $V(r)$.

If the Hilbert space of our system is that of $N_e$ particles bound to 0LL with angular momentum number not exceeding $N_{\phi}$ (disc geometry), to project our Hamiltonian means to compute the following matrix elements:

\begin{widetext}

\begin{equation}\label{pairi}
\left<0, n_1; 0, n_2 \right|\hat{V}\left|0,n_3; 0,n_4\right>=\sum_{m} V_m \sum_M \int d^2 z_1 d^2 z_2 \psi^{\ast}_{0,n_1;0,n_2}(z_1,z_2) \Psi_{M,m}(z_1,z_2)\times \int d^2 z_1 d^2 z_2 \psi^{\ast}_{0,n_3;0,n_4}(z_2,z_1) \Psi_{M,m}(z_1,z_2)
\end{equation}

{Here we defined $z_i=x_i+iy_i$,  and following \cite{Laughlin-83}, we use} explicit expressions for the wavefunctions in 0LL ,
\begin{eqnarray}
\Psi_{M,m}(z_1,z_2) = \frac{1}{2\pi l_B^2} \frac{1}{(2l_B)^{M+m}\sqrt{m! M!}} (z_1+z_2)^M (z_1-z_2)^m \exp \left(-\frac{|z_1|^2 + |z_2|^2}{4 l^2}\right) \\
\psi_{0,n_1;0,n_2}(z_1,z_2)= \frac{1}{2\pi l_B^{2}}\frac{1}{\sqrt{n_1! n_2! (2l_B^2)^{n_1+n_2}}} z_1^{n_1} z_2^{n_2} \exp\left(-\frac{|z_1|^2 + |z_2|^2}{4 l_B^2}\right)
\end{eqnarray}

we compute the scalar products separately

\begin{equation}
\int d^2 z_1 d^2 z_2 \psi^{\ast}_{0,n_1;0,n_2}(z_1,z_2) \Psi_{M,m}(z_1,z_2)  =  \frac{1}{\sqrt{2^{n_1+n_2}}} \sqrt{\frac{C_{n_1+n_2}^m}{C_{n_1+n_2}^{n_1}}}  \sum_{k_1+k_2 = n_1} (-1)^{k_2} C_{n_1+n_2-m}^{k_1} C_m^{k_2} \delta_{n_1+n_2,M+m}
\end{equation}

\end{widetext}

\noindent and pseudopotentials 

\begin{equation}\label{coor_frac}
V_m = \frac{1}{2^{2m+1}m!}\int_{0}^{\infty}d\varrho \varrho^{2m+1}V(l_B \varrho)e^{-\frac{\varrho^2}{4}}
\end{equation}

Though we use exact diagonalization, it is useful to deduce Hartree (direct) and Fock (exchange) terms, arising in the Hartree-Fock approximation of pair potential via Haldane pseudopotentials. These are usually given by an expression through Fourier transform of interaction \cite{Ank1} $g_H=\frac{\tilde{v}(q=0)}{2\pi l^2_B}$, $g_F=\frac{\int_{\mathbb{R}^2}d^2q \tilde{v}(q)e^{-\frac{q^2 l^2}{2}}}{(2\pi)^2}$. Using an alternative expression for Haldane pseudopotentials \cite{Haldane89}
\begin{equation}
    V_m=\int \frac{d^2 q}{(2\pi)^2}\tilde{v}(q) e^{-q^2 l_B^2} L_m(q^2)
\end{equation}
and well-known decompositions of exponent and Dirac delta in terms of Laguerre polynomials $\delta(x)=e^{-\frac{x}{2}}\sum_{k\ge 0} L_k(x),~e^{\frac{x}{2}}=2\sum_{k\ge 0} (-1)^k L_k(x)$ we have.

\begin{equation}\label{ACnDC}
g_{H}=2 l^2_B \sum_0^{\infty}  V_m,~g_{F} =  2 l^2_B \sum_0^{\infty} (-1)^m  V_m
\end{equation}

This connection provides us with some information on the nature of interaction by looking at the associated pseudopotentials. For example, if all are nonnegative, then the Hartree term is always greater than the Fock term; moreover, if the pseudopotentials are monotonous, the Fock term will be positive in itself. These statements will be useful for assessing the concrete pair potential we use. 

\section{Haldane pseudopotentials for Yukawa-like interaction}\label{pseudopo}

Apart from physical motivation given in the main text, Yukawa-like interactions $V_{\alpha}(r)= \frac{v_0 \alpha e^{-\alpha r}}{r}$ come with a purely computational advantage, as it is easy to compute respective Haldane pseudopotentials and observe the relevant qualitative features.

From (\ref{coor_frac}) we deduce, substituting $\varrho=2\alpha l_B \left(\sqrt{t+1}-1\right)$

\begin{widetext}

\begin{equation}
V_m=\frac{v_0 \alpha l_B \int_0^{\infty} d\varrho  \varrho^{2m} e^{-\alpha l_B \varrho}e^{-\frac{\varrho^2}{4}}}{2^{2m+1}m!} = \frac{v_0\alpha^2 l_B^2}{2^{2m+1}m!}\int_0^{\infty} \left(2\alpha l_B\right)^{2m}e^{-\alpha^2 l_B^2 t} \underbrace{\frac{\left(\sqrt{t+1}-1\right)^{2m}}{\sqrt{t+1}}}_{f(t)}dt 
\end{equation}
For the function on the right-hand side, the first $m$ derivatives $f^{(k)}(0)=0$ and $f^{(m)}(t)=\frac{\Gamma\left(m+\frac{1}{2}\right)}{\Gamma\left(\frac{1}{2}\right)}\frac{t^m}{\left(t+1\right)^{m+\frac{1}{2}}}$. Integrating by parts, we arrive at

\begin{equation}
V_m=\frac{\Gamma\left(m+\frac{1}{2}\right)}{2\Gamma\left(\frac{1}{2}\right)}\frac{v_0\alpha^2 l_B^2}{m!}\int_0^{\infty} e^{-\alpha^2 l_B^2 t} t^m \left(t+1\right)^{-m-\frac{1}{2}} dt=\frac{v_0\alpha^2 l_B^2}{2\Gamma\left(\frac{1}{2}\right)}\Gamma\left(m+\frac{1}{2}\right)U\left(m+1;\frac{3}{2};\alpha^2 l_B^2\right)
\end{equation}

\end{widetext}

\noindent here $U(k,l,a)$ is the confluent hypergeometry (Tricomi) function.

\begin{figure}[h]
    \centering
    \includegraphics[clip,trim=1.2cm 5cm 4.5cm 5cm,width=\linewidth]{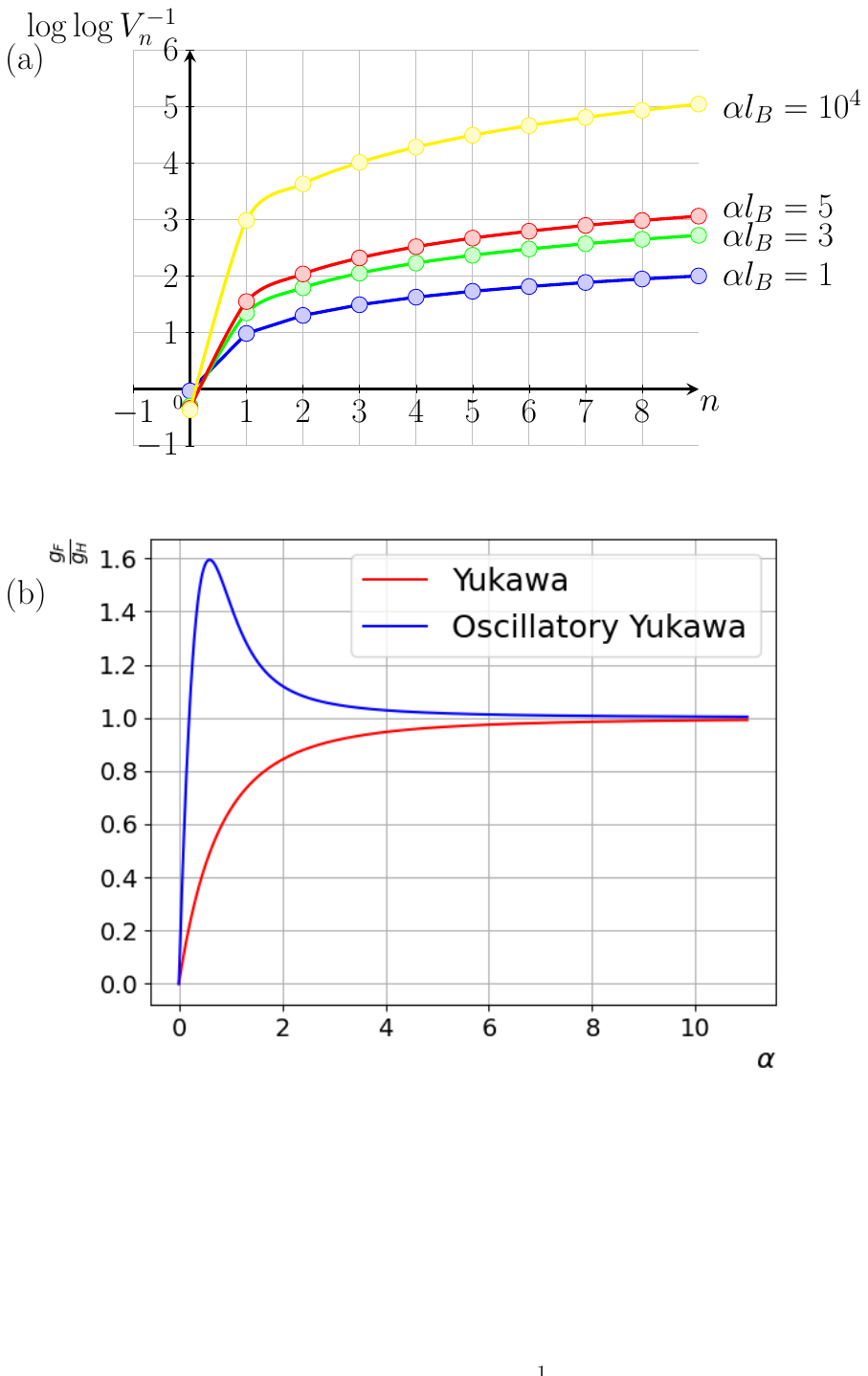}
    \caption{{(a)} Pseudopotentials (double logarithmic) for ranges addressed in this article. {(b) Fock and Hartree ratio for ``regular" (k=0) and oscillatory (k=2) Yukawa interactions}} 
    \label{un}
\end{figure}

As the range of interactions tend to zero, $\alpha\rightarrow \infty$, and asymptotic of Tricomi function gives us $U\left(m+1;\frac{3}{2};\alpha^2\right)\sim \frac{1}{\alpha^{2(m+1)}}$ \cite{AbrSteg}. Therefore, the ultra-short range limit indeed gives $2V_0=v_0,~V_{>0}=0$.

One important thing to note is that these pseudopotentials are nonnegative and monotonous. As we already said, this means that regardless of the value of parameters, the Hartree term will exceed the Fock term in our model. This may seem rather limiting, as new features occur mostly out of this area \cite{Coissard,Liu}. {Oscillatory Yukawa interaction $V_{k,\alpha}(r)=v_0 \alpha (k^2+1)\frac{e^{\alpha r} \cos k\alpha r}{r}$, however, lacks this drawback. With calculus very similar to ``regular" Yukawa, we arrive at the following expression for $V_m$, now nonmonotonous and shifting signs with number}

\begin{multline}
    V_m=\frac{v_0\alpha^2 (k^2+1) l_B^2 }{2\Gamma\left(\frac{1}{2}\right)}\Gamma\left(m+\frac{1}{2}\right)\times\\\Re\left({(1+ki)U\left(m+1;\frac{3}{2};(1+2i)^2\alpha^2 l_B^2\right)}\right)
\end{multline}

To illustrate the properties of pseudopotentials for parameters of choice for the systems we studied, we refer to \cref{un}. There are several key points we want to emphasize. {First, you can see that for a large region the ratio of Fock and Hartree is indeed in favor of the former for oscillatory Yukawa. Second, for the ranges considered in the paper, the bulk of interaction is indeed confined to a handful of pseudopotentials, here illustrated for Yukawa interaction}.  For our `effectively USR' interaction, even the first pseudopotentials are indeed at least 9 orders weaker than $V_0$. With the range increasing, $V_0$ starts to dwindle and $V_{>0}$ increases; as the range approaches the magnetic length, a change in pseudopotentials accelerates. Finally, you can see the asymptotics all the $V_n$ share regardless of the range. We chose to plot a double logarithm of $V_n$; due to variation in scale, it was the optimal way to showcase all those features in one chart, although not without drawbacks.

It should be pointed out that Haldane pseudopotentials for Yukawa were also treated for a different problem in \cite{AndrMoll}; the resulting expressions agree to our conclusions.

\section{Spherical geometry}\label{orb}

{Another conventional way to make the problem of interacting electrons in a magnetic field finite is to wrap the plane where they reside around a sphere; to maintain the field perpendicular to its surface one should place a magnetic monopole in its center \cite{Haldane83}. One-electron energy levels of this system, described in Refs. \onlinecite{WuYan0,WuYan1}, can be used to construct the Hilbert space. With respect to it, the matrix elements of an arbitrary (isotropic) pairwise interaction are expressed via combination of spherical Haldane pseudopotentials and Clebsch-Gordan coefficients. We briefly sum up the results given in \cite{Haldane83}

If the magnetic flux through the sphere is an integer number of flux quanta $N_\phi$, the dimensionless monopole strength, which we denote by $2Q$, has to be integer. The magnetic length $l_B$ corresponding to the magnetic field on the spherical surface is then proportional to its radius $R=\sqrt{|Q|}l_B$.

An electron confined to such a sphere has the following spectrum $E_n=\left(n+\frac{1}{2}+\frac{n(n+1)}{2|Q|}\right)\hbar \omega_C$, with each level being $2|Q|+2n+1$-fold degenerate. Due to it, the filling factor is somewhat ill-defined for a sphere.  being the ratio of a number of particles occuping the states to a number of states available
it now depends on the index of Landau level. The case of charge neutrality, which we are dealing with, is relatively easy. With monopole strength $2Q$ we have $4Q+2$ electrons occupying two full sublevels.

For a system of particles interacting via rotationally invariant pair potential $V(r)$ the second quantized Hamiltonian projected to the 0LL is then given by the same expression (\ref{H},\ref{h_unf}). The sum's indices now run $j_i=-Q,\ldots,Q$ and matrix elements now stand for}

\begin{equation}
V_{j_1j_2j_3j_4}=\sum_{L=0}^{2Q}\sum_{M=-L}^{L}\left<Qj_1,Qj_2\right|\left.LM\right>V_{2Q-L}\left<LM\right.\left|Qj_3,Qj_4\right>
\end{equation}
{where $\left<j_1m_1,j_2m_2\right|\left.JM\right> $ stand for Clebsch-Gordan coefficient, and $V_{2Q-L}$ are Haldane spherical pseudopotentials, i.e., they indicate the interaction energy of two particles with relative angular momentum $L$. A closed formula for $V_L$ is given in \onlinecite{WoMac} for an arbitrary potential $V(r)$. Being rotationally symmetric, the latter allows a multipole expansion $V(r)=\sum_{k}^{\infty}V_k \mathcal{P}_k(\cos \theta)$ (with polar angle $\theta$ and Legendgre polynomials $\mathcal{P}_k(z)$); in terms of its coefficients $V_k$ then }

\begin{equation}\label{cool_eq}
V^Q_L=  (-1)^L \sum_{k=0}^{2Q} V_k\left(2Q+1\right)^2 \left\{L~Q~Q\atop k~Q~Q\right\} \left(~Q~k~Q\atop -Q~0~Q\right)^2
\end{equation}
{where $\left\{j_1~j_2~j_3\atop j_4~j_5~j_6\right\}$ and $\left(j_1~~j_2~~j_3\atop m_1~m_2~m_3\right)$ are Wigner 6j- and 3j-symbols.
	
The only ingredient left one needs to perform ED in sphere geometry for the system in question is the multipole expansion for Coulomb and Yukawa interctions. The latter was computed in Ref. \onlinecite{YuMultip} $V_l=(2l+1) \alpha i_l(\alpha) k_l(\alpha)$, where $i_l$ and $k_l$ are \cite{AbrSteg} spherical Bessel functions of second and third kind respectively. As in case of disk, generalizing this expression to apply to oscillatory Yukawa can be performed, which gives us $V_l=5\Re(2l+1)(1+2i)\alpha i_l((1+2i)\alpha)k_l((1+2i)\alpha)$.}

\begin{figure*}
 \includegraphics[angle=270, width=\textwidth]{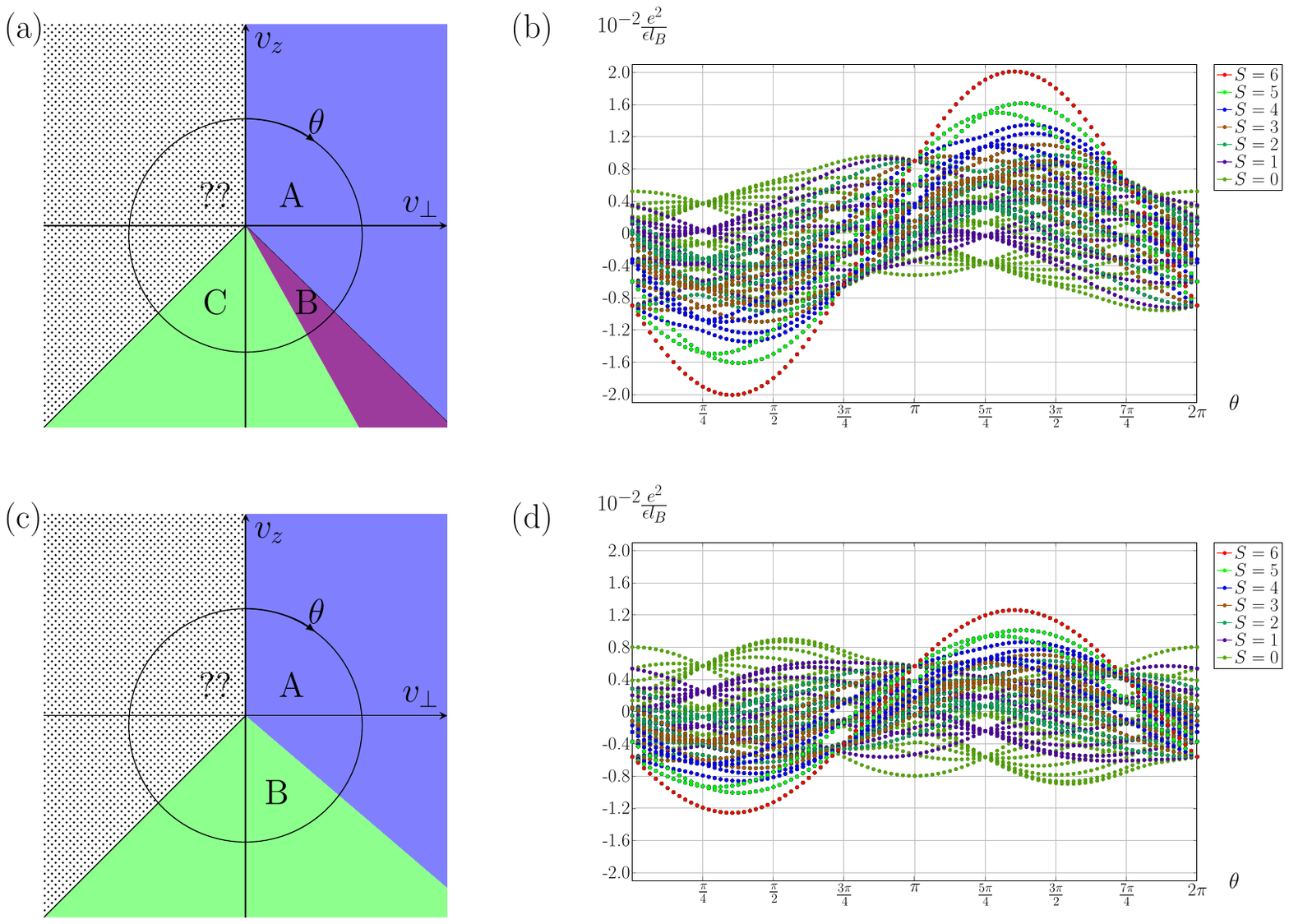}
	\caption{{Energy spectra for interacting electrons in graphene at charge neutrality on a sphere, $E_Z=0$, $2Q+1=6,~N_e=12$, for different valley symmetry breaking short range interactions, and proposed phase diagrams. (a,b) Oscillatory Yukawa, $\alpha^{-1}_z=\alpha^{-1}_{\perp}=l_B/2$ (c,d) Yukawa, $\alpha^{-1}_z=\alpha^{-1}_{\perp}=l_B/3$} }
    \label{kadurcharts}
\end{figure*}
    
{Using all of the above machinery, we computed the spectra for Yukawa weakly symmetry breaking interactions. The results are summed up in Fig. \ref{kadurcharts}. The description of phases coincides with the one we gave for disk geometry; the behavior of the system is qualitatively the same. There are however quantitative differences; curiously, it seems increasing range has stronger influence in case of a sphere. We perform some calculations that may explain this differences and assess the validity of both geometries in our system in the next section.}

\section{Comparison to mean field results}\label{dougandvladim}

{As we mentioned in the main text, one of the systems we considered (charge neutrality) has been recently researched \cite{Ank1} using Hartree-Fock methods. This gives us a benchmark for evaluating our results.

Qualitatively different phase diagrams in Ref. \onlinecite{Ank1} are grouped depending on whether Fock terms for intervalley interactions exceed Hartree terms or not. Formulae \ref{ACnDC} allow us to evaluate those for our model interactions; their dependencies on range are shown in Fig. \ref{un}, b.

Out of nine possible configurations $\left(\frac{g^{z,\perp}_F}{g^{z,\perp}_H}\lesseqgtr 1\right)$ the cases we considered encompass three. For the trivial (USR) case $\frac{g^{z,\perp}_F}{g^{z,\perp}_H}=1$; this configuration, according to Ref. \onlinecite{Ank1}, yield the same phase diagram as \cite{Kharitonov}, which agrees to our calculations. For Yukawa interaction we have $\frac{g^{\perp,z}_{F}}{g^{\perp,z}_{H}}<1$; Ref. \onlinecite{Ank1} predicts no new phases for configurations like this, yet the borders of existing phases along the line $v^{\perp}+v^{z}=0$ are shifted, with CDW sector dominating over FM -- again, in accord with our findings, -- and KD over AF. For oscillatory Yukawa all three options for $\frac{g_F}{g_H}$ are achievable; we, however, chose the ranges where $\frac{g^{\perp,z}_{F}}{g^{\perp,z}_{H}}>1$. In this case, Ref. \onlinecite{Ank1} predicts borders shifting in the opposite direction (with AF and CDW sector retreating from the transition lines to KD and FM states, respectively. Further, a new phase emerging on the border of FM and CDW is foreseen; this again agrees to our findings. Further still, another phase is found on the AF-KD border, but only in case of nonzero Zeeman, which we do not consider in this paper. However, the bandsctructures we calculated hint at a possible tower of states collapse in the thermodynamical limit, so whether a coexistence or an intermediary phase may be revealed here remains an open question.

Both FM and CDW phases being SSD states, more refined analysis of our numerical results can be carried out to assess their validity. In HF approach energies per flux quanta of these states can be expressed for FM as $-2g^{\perp}_{F}-g^z_{F}$ and for CDW as $2g^{z}_{H}-g^{z}_{F}$. As $g_H$ and $g_F$ depend linearly on interaction strengths $v_0^x=v_0 \sin\theta,~v_0^{z}=v_0 \cos\theta$, we can expect energies of FM and CDW obtained in ED to be trigonometric functions of $\theta$, namely $\frac{E_{FM}}{N_{\phi}}=-2g^{0}_{F}\sin\theta-g^{0}\cos\theta=-\sqrt{5}g^{0}_{F}\cos\left(\theta-\arccos\frac{1}{\sqrt{5}}\right)$, $\frac{E_{CDW}}{N_\phi}=(2g^0_{H}-g^0_{F})\cos\theta$. Therefore, we can obtain approximate values for Hartree and Fock couplings as fitting parameters for our numerical data.}

\begin{widetext}
\begin{center}
\begin{table}[H]
\centering
\begin{tabular}{||c | c | c | c||} 
 \hline
  Cases & $g_F, \cdot 10^{-3} \frac{e^2}{\epsilon l_B}$ & $2g_H-g_F, \cdot 10^{-3}\frac{e^2}{\epsilon l_B}$  & $\frac{g_F}{g_H}$ \\ [0.5ex] 
 \hline\hline
 Yukawa, $\alpha l_B=3$, disk geometry & 0.8590228 & 0.9716982 & 0.9384529 \\ 
 \hline
 Yukawa, $\alpha l_B=3$, sphere geometry & 0.9384862 & 1.33271 & 0.82642459 \\
 \hline
 Yukawa, $\alpha l_B=3$, actual values & 0.913771 & 1.086229 & 0.913771 \\
 \hline
 Oscillatory Yukawa, $\alpha l_B=2$, disk geometry & 1.0239947 & 0.86523 & 1.0840369 \\
 \hline
 Oscillatory Yukawa, $\alpha l_B=2$, sphere geometry & 1.497615 & 0.7231738 & 1.348723 \\ 
 \hline
 Oscillatory Yukawa, $\alpha l_B=2$, actual values & 1.1201688 & 0.879832 & 1.1201688 \\
 \hline
\end{tabular}
\caption{Fock and Hartree terms, derived from fitting FM and CDW state energies calculated by ED in various geometries for Yukawa-like potentials. Exact values are computed from \cref{ACnDC} for comparison.}
\label{tab:comparison}
\end{table}
\end{center}
\end{widetext}

{We see that the effects of range are undercounted in disk geometry and overcounted in sphere. The following simple calculation helps explain this discrepancy. In both geometries, we have similar expressions for energies of SSD states. Say, CDW state, $\Delta E_{CDW}=\left<CDW\right|H^{z}_{S-R}\left|CDW\right>=\sum_{i,j}2V^z_{ijji}-V^z_{ijij}$. In disk geometry, this sum is truncated at $i,j=N_{\phi}-1$, and we have for, say, the first (direct) term}

{
\begin{multline}
\Delta E_{CDW}^{d}=\sum_{i,j=0}^{N_{\phi}-1}2V^z_{ijji}=\\
=2\sum_{n=0}^{2(N_{\phi}-1)}V^z_n \sum_{k=0}^{2(N_{\phi}-1)} \sum_{i+j=k\atop i,j = 0}^{N_{\phi}-1} \left<0,i;0,j\right|\left.k-n,n\right>^2
\end{multline}}

{We broke the double sum of squared coefficients into subsums with fixed value of $i+j$; if the restriction $i,j<N_{\phi}$ is dropped, they are equal to 1 each due to orthogonality. However, because the momenta are truncated, some of them do not reach 1; thus, $ \frac{\Delta E^{d}_{CDW}}{N_{\phi}}$ is "almost" equal to $4\sum V^z_n=2g^z_H$, with higher pseudopotentials being increasingly undercounted. This causes the downplay of non-USR terms we see in disk geometry.

For the sphere, on the other hand}

{
\begin{multline}
\Delta E_{CDW}^{d}=\sum_{i,j=-Q}^{Q}2V^z_{ijji}=2\sum_{L=0}^{2Q}V^{Q,z}_{2Q-L} \times \\ \sum_{M=-L}^{L} \sum_{ i,j = -Q}^{Q} \left<Q,i;Q,j\right|\left.L,M\right>^2=2\sum_{L=0}^{2Q} V^{Q.z}_{2Q-L} (2L+1)
\end{multline}}

{For $ \frac{\Delta E^{d}_{CDW}}{2Q}$, again, the higher pseudopotentials are undercounted. However, for spherical pseudopotentials $V^Q_n$ there is no simple expression like (\ref{ACnDC}) expressing $g_{F,H}$; their divergence from disk $V_n$ apparently causes overestimation of effects of range we see. 

Qualitatively, though, both geometries yield similar results. For a short range interaction with limited number of pseudopotentials being relevant, as one can see from our estimates, the energies converge to correct results with increasing flux quanta numbr}

\begin{figure}
    \centering
    \includegraphics[width=\linewidth]{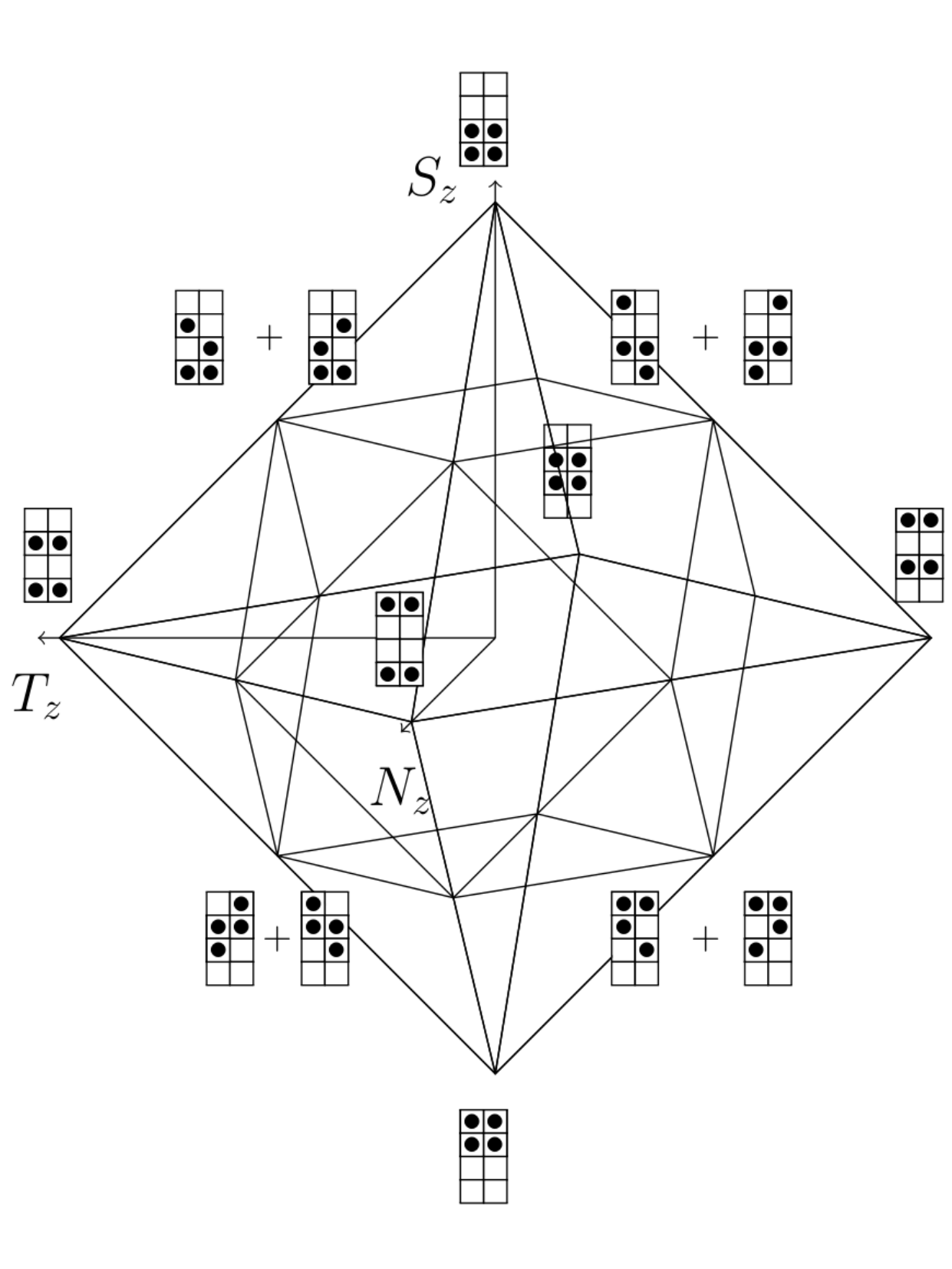}
    \caption{States of the (2,2,0) multiplet arranged as vertices of a polyhedral graph in $(T_z,N_z,S_z)$ space.
    }
    \label{convex}
\end{figure}

\section{SU(4) multiplets of states}\label{trunca}

To fully describe a state of our system, we use the Fock space of $N_e$ fermions with spin and valley spin $\frac{1}{2}$, each of them having an angular momentum $n\hbar,~n = 0, \ldots, N_{\phi}-1$. This space delivers a $C_{4 N_{\phi}}^{N_e}$ dimensional representation of $SU(4)$, which breaks into a direct sum of irreducible representations, or multiplets. Below, we sum up those statements regarding $SU(4)$ representation theory that are most relevant to the studies of multielectron states in graphene; we also choose the language so that their connection to physics is as transparent as possible.

Firstly, we adopt, following ref. \onlinecite{FischerDzRom}, the nuclear physics notation for $SU(4)$ spin configurations, denoting $\downarrow K',\downarrow K, \uparrow K', \uparrow K$ as down, up, strange, and charmed ($d,u,s,c$), respectively. This way, we can use it both for the $\mathbb{C}^4$ space of one electron $SU(4)$ polarizations, where it acts tautologically and for the many-electron spaces of states.

Like $su(2)$, the multiplets of $su(4)$ coincide with multiplets of a corresponding special linear algebra; i.e. we may study the generators of $sl(4,\mathbb{C})$ instead. Among those, there are 3 traceless diagonal matrices and 12 upper/lower-diagonal ones. We choose the base in such a way that the 3 diagonal generators correspond to spin, valley spin and N\'{e}el vector polarizations ($2S_z=C_{dd}+C_{uu}-C_{ss}-C_{cc},~2T_z=C_{dd}-C_{uu}+C_{ss}-C_{cc},~2N_z=C_{dd}-C_{uu}-C_{ss}+C_{cc}$), and the rest are ``matrix unities" $C_{km}$. The commutation relations are deduced from the matrix unities ones $[C_{ij},C_{kl}]=\delta_{il}C_{jk}-\delta_{jk}C_{il}$.

Now, for a multiplet, a base can be chosen so it is an eigenbase for each of (commuting) polarization operators, so $S_z,T_z, N_z$ are good quantum numbers. Out of the remaining 12 operators, for each polarization, four will commute with its operator, leaving the quantum number unchanged; four will increase it by 1 and four will decrease it. If we represent each vector by the triplet of its quantum numbers, it will span a convex symmetric polyhedron, with its vertices standing for ``highest weight" vectors. Choosing the vector such that $S_z\ge T_z\ge N_z$, we can define a Young diagram with three rows consisting of $L_1=S_z+T_z,~L_2=S_z-N_z,~L_3=T_z-N_z$ cells; such a diagram defines an irrep uniquely. If we take $p_1=L_1-L_2=T_z+N_z,~p_2=L_2-L_3=S_z-T_z,~p_3=L_3=T_z-N_z$,the dimension of a representation is given by the formula \cite*{Pfeifer,FischerDzRom}
\begin{widetext}
\begin{equation}\label{mt}
    d(p_1,p_2,p_3)=\frac{1}{2!3!}(p_1+1)(p_2+1)(p_3+1)(p_1+p_2+2)(p_2+p_3+2)(p_1+p_2+p_3+3)
\end{equation}
\end{widetext}

In the case of $SU(2)$, it is easy to deduce to what multiplet an $S_z$ eigenvector belongs -- we should either compute the (quadratic Casimir) operator $S^2$ action on them or act with step operators $S_{\pm}$ until it reaches the highest weight vector. Similarly to these strategies, it is possible to classify the vectors by action of three Casimir operators \cite{Pais}, or act until we reach the ``corner" of our irrep -- the highest weight vector. In practice, however, we judge by the action of unterperturbed Hamiltonian (which can in principle be expressed in terms of Casimirs), and then check the dimension of our eigenspace against (\ref{mt}) to make sure that there is no further degeneracy of several multiplets.

In terms of fermionic Fock operators in the central Landau level, the set of aforementioned matrix unities looks like \cite{FischerDzRom}

\begin{equation}
    \Hat{C}_{i,j}=\sum_{m=0}^{N_{\phi}-1} \hat{c}^{\dagger}_{m,i}\hat{c}_{m,j}
\end{equation}
where $i,j$ are any of 16 combinations of $d,u,s,c$. Notice that these operators commute with total angular momentum.
This means that to study a certain $su(4)$ multiplet, you may restrict yourself to a certain angular momentum sector, thus greatly reducing the number of calculations needed.

To make all of the above more tangible, we can consider a concrete case of charge neutrality with $N_{\phi}=2,N_e=4$. This means a relevant Young diagram is $\ytableausetup{boxsize=0.15cm}{\ydiagram{2,2}}$.  Placing the vectors of such a multiplet in a three-dimensional space spanned by quantum numbers $S_z,T_z,N_z$, all the highest weight vectors will form a set of vertices for the octahedron, their descendants under the action of ``ladder operators" will occupy the middle points of octahedral edges, and finally, their two{{-}}second descendants will dwell in the origin. The states are shown by their flavour content. A part of these states are shown in \cref{convex}; one can imagine algebra generators acting along certain edges of the octahedron.

\bibliography{main}
\end{document}